%% file: main.tex
\documentclass[conference]{IEEEtran}
\usepackage{geometry}
\AtBeginDocument{
  \setlength{\abovedisplayskip}{3pt plus 1pt}
  \setlength{\belowdisplayskip}{3pt plus 1pt}
}
\IEEEoverridecommandlockouts
\usepackage{orcidlink}
\usepackage{cite}
\usepackage{amsmath,amssymb,amsfonts}
\usepackage{algorithmic}
\usepackage{graphicx}
\usepackage{textcomp}
\usepackage{xcolor}
\def\BibTeX{{\rm B\kern-.05em{\sc i\kern-.025em b}\kern-.08em
    T\kern-.1667em\lower.7ex\hbox{E}\kern-.125emX}}
\usepackage{tabularx}    
\usepackage[T1]{fontenc}
\usepackage[utf8]{inputenc}
\usepackage{booktabs}
\usepackage{xspace}
\usepackage{url}
\usepackage{enumitem}
\usepackage{array}
\usepackage{multirow}
\usepackage{tikz}
\usetikzlibrary{arrows.meta, positioning, shapes.geometric, calc, backgrounds, fit}

\usepackage{tikz}
\usepackage{hyperref}

\newcommand{\orcidID}[1]{\href{https://orcid.org/#1}{\includegraphics[width=9pt]{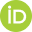}}}

\begin{document}
\title{Autonomy in Check: Governor-Mediated Adaptive Security at the Edge}

% \thanks{This research is supported by the Finnish Doctoral Program Network in Artificial Intelligence, AI-DOC (decision number VN/3137/2024-OKM-6) and Research Council of Finland funded projects 6G Flagship (369116) and Profi6 (336449).}

% \author{\IEEEauthorblockN{1\textsuperscript{st} Ijaz Ahmad}
% \IEEEauthorblockA{\textit{Centre for Wireless Communications} \\
% \textit{University of Oulu}\\
% Oulu, Finland \\
% ahmad.ijaz@oulu.fi}
% \and
% \IEEEauthorblockN{2\textsuperscript{nd} Ijaz Ahmad}
% \IEEEauthorblockA{\textit{VTT Technical Research} \\
% \textit{Centre of Finland}\\
% Espoo, Finland \\
% ijaz.ahmad@vtt.fi}
% \and
% \IEEEauthorblockN{3\textsuperscript{rd} Flavio Esposito}
% \IEEEauthorblockA{\textit{Department of Computer Science} \\
% \textit{Saint Louis University}\\
% Saint Louis, Missouri, USA \\
% flavio.esposito@slu.edu}
% \and
% \IEEEauthorblockN{4\textsuperscript{th} Erkki Harjula}
% \IEEEauthorblockA{\textit{Centre for Wireless Communications} \\
% \textit{University of Oulu}\\
% Oulu, Finland \\
% erkki.harjula@oulu.fi}
% }

% \author{\IEEEauthorblockN{Anonymous Authors}}

\author{
\IEEEauthorblockN{
Ijaz Ahmad\IEEEauthorrefmark{1}\orcidID{0000-0002-6152-8947},
Ijaz Ahmad\IEEEauthorrefmark{2}\orcidID{0000-0003-1101-8698},
Flavio Esposito\IEEEauthorrefmark{3}\orcidID{0000-0002-7798-4584},
and Erkki Harjula\IEEEauthorrefmark{1}\orcidID{0000-0001-5331-209X}
}
\IEEEauthorblockA{\IEEEauthorrefmark{1}
Centre for Wireless Communications, University of Oulu, Oulu, Finland\\
Email: ahmad.ijaz@oulu.fi, erkki.harjula@oulu.fi}
\IEEEauthorblockA{\IEEEauthorrefmark{2}
VTT Technical Research Centre of Finland, Espoo, Finland\\
Email: ijaz.ahmad@vtt.fi}
\IEEEauthorblockA{\IEEEauthorrefmark{3}
Department of Computer Science, Saint Louis University, St. Louis, Missouri, USA\\
Email: flavio.esposito@slu.edu}
}

\maketitle

\begin{abstract}
Adaptive security at the network edge increasingly relies on automated planners, including rule-based controllers, learned policies, and LLM-assisted agents, that translate observations into enforcement actions. Once such a planner can influence live policy state, syntactic validity is not enough. A semantically wrong action, produced from incomplete or manipulated observations, can be faithfully executed by an enforcement substrate that cannot judge mission context.
We address this problem by treating the boundary between planner output and kernel enforcement input as the primary security object. We propose a split-control architecture in which an untrusted planner emits typed security intents, a deterministic governor checks each intent against safety, resource, temporal-stability, and proportionality invariants, and only admitted actions are bound to signed receipts and compiled into pre-installed eBPF map updates. The paper formalizes this trust-boundary problem, defines three threat classes, develops the governor admission predicate, and reports an end-to-end prototype.
Across rule-based and LLM-assisted planners on a Raspberry Pi 5 testbed connected to the university's 5G Test Network, the governor admits, rejects, and bounds intents at microsecond cost without disrupting protected-flow regularity. The contribution is conceptual as much as empirical: adaptive security does not need to trust the author of an action. It needs a mediation boundary that decides whether the action is admissible.
\end{abstract}


\section{Introduction}
\label{sec:intro}

Modern edge systems increasingly rely on adaptive security mechanisms that adjust enforcement policies in response to changing network conditions, resource constraints, and observed threats. These mechanisms are often driven by automated planners, ranging from rule-based controllers to learning-based and LLM-assisted agents, that translate observations into actions such as monitoring, rate limiting, tier escalation, or isolation~\cite{ahmad2025adapt,sangiorgi2025deauth}. This shift improves responsiveness, but it also changes the security problem. The adaptation process itself becomes part of the attack surface.

Current adaptive security systems often trust planner outputs once they are syntactically valid, even though semantically unsafe actions can emerge from manipulated observations or imperfect reasoning. A planner operating on incomplete, noisy, or adversarially shaped inputs may issue an action that is locally consistent with policy but unsafe for the mission context. The enforcement substrate then executes that action faithfully. The risk is not only that a threat is missed. It is also that a controller may throttle protected traffic, under-react to an active attack, or oscillate rapidly between enforcement states because its inputs have been shaped by an adversary~\cite{greshake2023indirect,nist2026rfi,nist2026monitoring}.

This problem is not tied to a single application domain. In industrial control systems, security mechanisms must account for control-loop timing and survivability requirements~\cite{cardenas2008control}. In vehicular and emergency-response edges, adaptation must support delay-critical decisions under changing mobility and load~\cite{liu2021vec}. Private 5G deployments face similar pressure because mixed-criticality traffic classes share the same access layer~\cite{fue2025private5g}. O-RAN xApp/rApp ecosystems add another instance of the same problem, since control-plane intelligence can influence operational state and may itself be compromised~\cite{oran_wg11_threat}.

Healthcare ward networks provide a concrete example of the same pattern. Clinical alarms and routine telemetry may traverse the same access network~\cite{jointcommission2013alarm ,sendelbach2013alarm}. An adaptive controller that is too restrictive can delay or suppress urgent flows, while one that is too permissive leaves the same path open to a compromised endpoint. Across these domains, the key question is no longer only how to detect malicious behavior, but how to constrain what an adaptive controller is allowed to do when its inputs cannot be fully trusted.

Existing closed-loop and intent-based control frameworks automate policy refinement and actuation, while programmable substrates such as extended Berkeley Packet Filter (eBPF) provide efficient in-kernel observability and enforcement~\cite{etsi_zsm009,rfc9315,soldani2023ebpf,ahmad2026adaptive,cilium2026intro}. These foundations are important, but efficiency and well-typed actuation do not by themselves establish semantic safety. The kernel can enforce a rate limit or update a policy map, but it cannot decide whether the requested update is appropriate for the current mission context. Likewise, model-level defenses for LLM and agent security reduce planner-side risk, but they do not remove the need to mediate unsafe outputs that may still be produced~\cite{chatzimiltis2025agentic,sharma2025mobillm}.

This paper treats the planner-enforcement boundary as the primary security object. We propose a \emph{split-control} architecture grounded in the reference-monitor principle~\cite{anderson1972computer}. The planner may use rules, learned logic, or LLM-assisted reasoning, but it cannot directly modify enforcement state. Instead, it emits typed security intents. A deterministic governor mediates each intent against safety, resource, temporal-stability, and proportionality invariants before any action can be compiled into the eBPF enforcement substrate.

\noindent This paper makes the following contributions:
\begin{itemize}[leftmargin=*,topsep=2pt,itemsep=1pt]
  \item \textbf{Trust-boundary formulation.} We frame adaptive edge security as a trust-boundary problem in which policy-valid actions can be semantically unsafe under manipulated observations or imperfect reasoning. We define three concrete threat classes at the planner-enforcement interface: context manipulation, policy oscillation, and residual-window exploitation (Section~\ref{sec:model}).

  \item \textbf{Governor-mediated architecture.} We design a deterministic admission predicate over a restricted intent vocabulary. The predicate checks syntactic validity, protected-flow safety, resource headroom, temporal stability, and proportionality before enforcement. This separates intent mediation from planner-specific mechanisms such as action masking in safe reinforcement learning or per-call runtime verification (Section~\ref{sec:governor}).

  \item \textbf{Bounded enforcement with integrity.} We bind every admitted intent to an HMAC-protected receipt and confine the compiler to a pre-installed set of enforcement state space, so neither the planner nor a compromised compiler can synthesize new policy logic. Every approved action lands in an append-only audit log (Section~\ref{sec:governor}).

  \item \textbf{Prototype evaluation under structured threats.} We implement the design on a Raspberry~Pi~5 attached to the University's 5G Test Network with three ESP32 endpoints. We evaluate rule-based and LLM-assisted planners under the three threat classes, characterize governor overhead, and compare governed behavior against static and no-governor baselines (Section~\ref{sec:eval}).
\end{itemize}

The results support a simple design principle: autonomy belongs in planning, but actuation should remain mediated, bounded, and auditable. Section~\ref{sec:related-work} situates the work against closed-loop control, eBPF enforcement, AI-driven control ecosystems, LLM security, and runtime policy mediation. Section~\ref{sec:model} develops the system and threat model. Section~\ref{sec:governor} presents the governor-centered design. Section~\ref{sec:proto} describes the prototype and attack emulation methodology. Section~\ref{sec:eval} reports the evaluation results and discusses deployment implications.

\section{Related Work}
\label{sec:related-work}

\noindent\textbf{Closed-loop and intent-based control.}
Closed-loop network control has developed through autonomic networking, zero-touch service management, intent-based networking, and edge-cloud orchestration. ETSI ZSM models closed-loop automation as an observe, decide, and actuate process with minimal human intervention~\cite{etsi_zsm009}. Intent-based networking similarly raises the interface from low-level configuration to desired outcomes, leaving the system to refine those outcomes into device-level actions~\cite{rfc9315}. Recent work on adaptive trust and edge-cloud orchestration applies related ideas to 6G and IoT settings, where service quality, resource headroom, and security posture must be adjusted at runtime~\cite{ahmad2025adapt,ahmad24adaptive}. These systems are important foundations for adaptive security, but they typically treat the controller, planner, or intent source as trusted once its output is well formed. Our work focuses on the complementary problem. A planner output may be syntactically valid and still semantically unsafe when observations are manipulated or reasoning is imperfect.

\noindent\textbf{Programmable enforcement with eBPF.}
eBPF has become a practical substrate for programmable observability, traffic control, and cloud-native security because it offers stable kernel hooks, verifier-checked programs, and low-overhead access to packet and system events~\cite{soldani2023ebpf,cilium2026intro}. Recent systems also use eBPF to observe AI and agent workloads at the system level~\cite{zheng2025agentsight}. At the same time, work on BPF hardening and kernel-extension safety shows that verifier acceptance and efficient execution are not the same as end-to-end security correctness~\cite{jin2024beebox,sun2025aee,jia2025rex}. eBPF can enforce a rate limit, select an inspection path, or update a policy map efficiently. It does not decide whether the requested action is appropriate for the mission context. 

\noindent\textbf{Open and AI-driven control ecosystems.}
Open and programmable control ecosystems make the planner--enforcement boundary more important. O-RAN introduces xApps and rApps that can optimize radio and service behavior through near-real-time and non-real-time control loops, but this openness also creates risks from compromised or malicious control functions~\cite{oran_wg11_threat,oRanSecurity2025}. Recent agentic-AI proposals extend this direction by using LLM-assisted reasoning for closed-loop RAN or network-security tasks~\cite{chatzimiltis2025agentic,sharma2025mobillm}. LLM-integrated systems also face prompt injection, indirect prompt injection, and adversarially shaped inputs that can cause an agent to infer or follow unintended context~\cite{greshake2023indirect}. Guidance on deployed AI systems further emphasizes that monitoring remains difficult once AI components interact with tools, external data, and operational environments~\cite{nist2026rfi,nist2026monitoring}. These works reduce planner-side risk, but they do not remove the need to mediate unsafe outputs that may still be produced.

\noindent\textbf{Runtime safety and policy mediation.}
The closest conceptual anchor for our work is runtime policy mediation. The reference-monitor principle requires security-sensitive operations to be mediated by a mechanism that is always invoked, tamper-resistant, and small enough to reason about~\cite{anderson1972computer}. Schneider's work on enforceable security policies formalizes policies that can be enforced by monitoring execution and suppressing disallowed actions~\cite{schneider2000enforceable}. Runtime verification checks execution traces against formal specifications while a system runs~\cite{sanchez2019runtime}. In learning-enabled control, shielding prevents unsafe actions from a learner before they affect the environment~\cite{alshiekh2018shielding}. Our governor follows the same safety-oriented spirit, but it mediates a different object. A shield typically masks individual learner actions against a verified safety model. Runtime verification observes program traces against a specification. Intent-based frameworks refine desired outcomes into actions while usually assuming that the intent source is trusted. Our governor instead mediates typed security intents before they can modify live enforcement state. The planner is explicitly outside the trusted computing base, and each proposed intent must clear the predicate $\Gamma$ over deployment-specific invariants such as $\mathcal{F}_{\mathrm{crit}}$, $r_{\min}$, $\Delta_{\min}$, and $\phi$.

\noindent\textbf{Identified gap.}
Prior work has advanced adaptive control, programmable enforcement, open control ecosystems, AI-assisted security, and runtime safety. These lines of work make critical edge systems more flexible, but they often assume that a well-formed controller output is safe to execute. We take a different position. In critical edge deployments, the planner's output is part of the attack surface. A semantically wrong but syntactically valid action can delay protected traffic, weaken containment, or drive unstable policy changes. Our architecture therefore separates planning from actuation: the planner proposes, the governor mediates, and the eBPF substrate enforces only actions that pass explicit safety and stability checks.

\section{Trust Boundary and Threat Model}
\label{sec:model}

\subsection{Deployment Context and System Model}
We consider a critical edge network carrying mixed-criticality traffic over a 5G access layer, with processing tiers at the local edge, MEC, and the cloud. Some flows are urgent and delay-sensitive, while others are routine and tolerant of inspection or queuing delays. A security controller observes the communication state and decides whether to continue passive monitoring, rate-limit a traffic class, escalate inspection to a higher-capacity tier, or isolate a flow subset. Each decision affects both security posture and the service quality of the flows under management.

The proposed defense mechanism is structured as three logical modules. A \emph{planner} receives the current observation state and proposes a typed security intent. A \emph{governor} checks whether that intent is admissible under the current mission and resource constraints. A \emph{compiler} translates approved intents into bounded updates over a pre-installed eBPF enforcement substrate. The planner may use rule-based logic, a learned policy, or an agentic reasoning component. Its internal mechanism is outside the trust boundary argument. What matters is that it cannot modify enforcement state without governor approval.

\subsection{Threat Model}
\label{sec:threats}

The governor and enforcement substrate are part of the trusted computing base, whereas the planner is not. The planner therefore cannot load arbitrary eBPF programs, hold file descriptors to enforcement maps, or directly invoke the compiler. This boundary is what we study, not something we abstract away. The adversary may inject malicious traffic~\cite{gimhana2025mmtc}, compromise a non-critical endpoint device, or influence the telemetry and context presented to the planner. From these capabilities, three threat classes emerge, summarized in Table~\ref{tab:threats} and illustrated in Fig.~\ref{fig:threats}.

\noindent\textbf{T1~-- Context manipulation.} The adversary manipulates observable inputs, such as inflated anomaly signals, masked malicious flows, or workloads shaped to mislead, to induce an action that is syntactically valid but semantically wrong. Suppressing the measurement of an ongoing attack, for instance, causes the planner to perceive low threat and propose passive monitoring when isolation is warranted. This class maps directly onto indirect injection attacks studied in LLM-integrated systems~\cite{greshake2023indirect}, where the attack targets the input environment rather than the model itself.

\noindent\textbf{T2~-- Policy oscillation.} Rather than inducing a single large policy error, the adversary alternates observable behavior at a rate $f_{\mathrm{adv}}$ chosen to keep the planner near a decision boundary. If $f_{\mathrm{adv}} > 1/\Delta_{\min}$, where $\Delta_{\min}$ is the governor's minimum inter-action cooldown, each enforcement transition creates a brief window during which the active policy is mismatched with the actual traffic condition. Repeated over time, this degrades both protection coverage and service quality without requiring any single catastrophic decision.

\noindent\textbf{T3~-- Residual-window exploitation.} Every enforcement state transition involves a brief window before the new policy takes full stable effect. An adversary who can trigger policy transitions, whether through context manipulation or oscillation, can time traffic delivery to coincide with this window. The primary mitigation is bounding the time from governor approval to stable kernel enforcement, addressed in Section~\ref{sec:integrity}.

\begin{table}[htpb]
  \centering
  \caption{Threat Classes at the Planner--Governor Interface}
  \label{tab:threats}
  \renewcommand{\arraystretch}{1.25}
  \small
  \begin{tabular}{p{0.5cm} p{3.8cm} p{3.2cm}}
    \toprule
    \textbf{Class} & \textbf{Adversary action} & \textbf{Governor defense} \\
    \midrule
    T1 & Shape observations to deceive the planner & Proportionality invariant (I5); uncertainty gating \\[3pt]
    T2 & Alternate behavior to drive rapid enforcement switching & Temporal stability invariant (I4) \\[3pt]
    T3 & Time delivery to coincide with transition gaps & Bounded compiler latency; receipt sequencing \\
    \bottomrule
  \end{tabular}
  \vspace{-20pt}
\end{table}

\input{figs/fig1}

\subsection{Problem Formulation: Governor-Mediated Actuation}
\label{sec:formulation}

To formalize how the governor constrains planner actions under the threats described above, we model each decision as an intent proposed from the current system state and independently admitted before enforcement. In particular, at decision epoch $t$, the security controller observes
\begin{equation}
  \mathbf{x}_t =
  [\mathbf{o}_t,\, \mathbf{c}_t,\, \mathbf{r}_t,\,
   \mathbf{q}_t,\, \mathbf{h}_t,\, \mathbf{m}_t,\, u_t],
  \label{eq:state}
\end{equation}
where $\mathbf{o}_t$ denotes traffic observations, $\mathbf{c}_t$ flow criticality labels, $\mathbf{r}_t$ resource headroom per enforcement tier, $\mathbf{q}_t$ queue and path state, $\mathbf{h}_t$ a bounded history of recent policy decisions, $\mathbf{m}_t$ mission-context indicators such as active alarm sessions and protected service classes, and $u_t\in[0,1]$ an uncertainty signal derived from measurement freshness and cross-source consistency.

Let $\mathcal{A}$ denote the typed intent vocabulary and let $\mathcal{L}=\{\text{local},\,\text{MEC},\,\text{cloud}\}$ denote the available execution tiers. A planner proposes an intent $a_t\in\mathcal{A}$ and a tier $\ell_t\in\mathcal{L}$, which can be abstracted as the following planner-side problem:
\begin{equation}
\label{eq:objective}
\begin{aligned}
\underset{
  \substack{
    a_t \in \mathcal{A}\\
    \ell_t \in \mathcal{L}
  }
}{\operatorname{min.}}
\quad
& R_t(a_t,\ell_t)
  + \lambda C_t(a_t,\ell_t)
  + \mu O_t(a_t,\mathbf{h}_t) \\
\mathrm{s.t.}
\quad
& D_t^{\mathrm{prot}} \le D_{\max}, \\
& U_t \le U_{\max}, \\
& a_t \in \mathcal{A}_{\mathrm{safe}}(\mathbf{x}_t).
\end{aligned}
\end{equation}
Here, $R_t$ denotes the planner's estimate of residual security risk, $C_t$ denotes resource and coordination cost, and $O_t$ penalizes unstable policy changes such as reversals on the same scope. $D_t^{\mathrm{prot}}$ is the protected-flow service bound and $U_t$ is a resource-utilization bound. The state $\mathbf{x}_t$ is supplied by the state builder at each epoch.
The safe action set $\mathcal{A}_{\mathrm{safe}}(\mathbf{x}_t)$ is not computed by the planner. It is the output of the governor admission predicate, evaluated independently of how the proposed intent was generated. A planner operating on manipulated observations may still produce an intent in $\mathcal{A}$; the governor determines whether that intent belongs in $\mathcal{A}_{\mathrm{safe}}(\mathbf{x}_t)$ under the actual system state. This separation between proposal and admissibility is the central structural property of the design.

\noindent\textbf{Prototype instantiation.}
Equation~\eqref{eq:objective} is a conceptual abstraction of the planner-side objective. The prototype does not implement a numerical optimizer for $R_t$, $C_t$, or $O_t$, and the governor does not trust the planner's estimates of these terms. Instead, the governor evaluates only the invariant predicate in Section~\ref{sec:predicate}. The state builder provides the a bounded severity signal from smoothed drop and queue-pressure indicators, the uncertainty signal $u_t$, and the intent impact ordinal shown in Table~\ref{tab:intents}. The proportionality bound is $\phi(s,u)=\max(1,\,4s\cdot c(u))$, where $c(u)=1$ for $u\le0.5$ and decreases linearly to $0$ at $u=1$. The floor of $1$ keeps \textsc{Rollback} admissible at any uncertainty. The cooldown $\Delta_{\min}$ and reserve $r_{\min}$ instantiate the temporal-stability and protected-headroom checks used by I4 and I3.

\section{Governor-Centered Design}
\label{sec:governor}

\subsection{Bounded Intent Vocabulary}
\label{sec:vocab}

Restricting the planner to a fixed vocabulary is the first step in bounding its actuation authority. Rather than accepting free-form policy expressions or arbitrary kernel operations, the governor sees only the typed intents in Table~\ref{tab:intents}. The vocabulary covers passive observation, bounded rate control, tier escalation, non-critical isolation, human-approval requests, and state rollback. High-impact actions are named explicitly so the governor can apply proportionality checks independently of planner intent.

\begin{table}[htpb]
  \centering
  \vspace{-10pt}
  \caption{Typed security intent vocabulary with impact ordinal.}
  \label{tab:intents}
  \renewcommand{\arraystretch}{1.18}
  \footnotesize
  \setlength{\tabcolsep}{3pt}
  \begin{tabular}{@{}p{1.95cm}p{5.95cm}@{}}
    \toprule
    \textbf{Intent (ord)} & \textbf{Parameters and effect} \\
    \midrule
    \textsc{Monitor} (0)
      & scope, duration; passive observation, no traffic effect \\[2pt]
    \textsc{RateLimit} (2)
      & scope, rate tier, duration; bounded throttling with mandatory exemptions for flows in $\mathcal{F}_{\mathrm{crit}}$ \\[2pt]
    \textsc{Escalate} (3)
      & scope, target tier, timeout; requests higher-capacity inspection without modifying the local rate policy \\[2pt]
    \textsc{Isolate-NC} (4)
      & scope, duration; isolates only non-critical flows or devices; critical flows are unaffected by construction \\[2pt]
    \textsc{ReqApproval} (0)
      & reason, suggested action, urgency; suspends direct actuation and requests operator review \\[2pt]
    \textsc{Rollback} (1)
      & receipt-id, target state; reverts to the last governor-approved stable enforcement state \\
    \bottomrule
  \end{tabular}
  \vspace{-10pt}
\end{table}

\subsection{Admission Predicate}
\label{sec:predicate}

The governor admits $a_t$ if and only if $\Gamma(a_t, \mathbf{x}_t) = 1$, where $\Gamma = I_1 \wedge I_2 \wedge I_3 \wedge I_4 \wedge I_5$. The safe action set follows as $\mathcal{A}_{\mathrm{safe}}(\mathbf{x}_t) = \{a \in \mathcal{A} \mid \Gamma(a,\mathbf{x}_t)=1\}$. Each invariant addresses a distinct failure mode.

\textbf{I1~-- Syntactic validity.} $a_t$ belongs to the declared vocabulary with valid parameter types and in-range values. Syntactic validity is necessary but not sufficient; the remaining invariants check semantic correctness.

\textbf{I2~-- Flow safety.} A restrictive action may not affect flows in $\mathcal{F}_{\mathrm{crit}}$ without an explicit policy exemption:
\begin{equation}
\label{eq:flow-safety}
\begin{aligned}
&\mathrm{scope}(a_t)\cap \mathcal{F}_{\mathrm{crit}}=\emptyset \\
&\quad \vee\ 
\mathrm{type}(a_t)\in
\{\textsc{ReqApproval},\textsc{Rollback}\}.
\end{aligned}
\end{equation}
Regardless of what the planner proposes, this invariant prevents urgent flows from being throttled or isolated without explicit authorization.

\textbf{I3~-- Resource bound.} Applying $a_t$ must leave sufficient bandwidth headroom for protected services:
\begin{equation}
  \mathrm{headroom}(\mathbf{r}_t, a_t) \geq r_{\min}.
  \label{eq:resource}
\end{equation}

\textbf{I4~-- Temporal stability.} The last approved action on $\mathrm{scope}(a_t)$ must have occurred at least $\Delta_{\min}$ epochs prior:
\begin{equation}
  t - t_{\mathrm{last}}\!\bigl(\mathrm{scope}(a_t)\bigr) \geq \Delta_{\min}.
  \label{eq:temporal}
\end{equation}
An adversary driving oscillation at $f_{\mathrm{adv}} > 1/\Delta_{\min}$ cannot force enforcement changes at that rate. The governor rejects each intent until the cooldown elapses, bounding the achievable oscillation rate above by $1/\Delta_{\min}$ regardless of how quickly the planner proposes new intents.

\textbf{I5~-- Proportionality.} The impact of $a_t$ must not exceed what observed threat severity and current uncertainty jointly support:
\begin{equation}
  \mathrm{impact}(a_t) \leq \phi\!\bigl(\mathrm{severity}(\mathbf{o}_t),\, u_t\bigr),
  \label{eq:proportional}
\end{equation}
where $\phi$ decreases monotonically in $u_t$. As uncertainty rises, $\phi$ contracts toward Low, directing the planner to \textsc{Monitor}, \textsc{ReqApproval}, or \textsc{Rollback} rather than direct mitigation.

On rejection, the governor returns a typed reason identifying which invariant failed and, where applicable, a lower-impact alternative. This allows the planner to revise its proposal without a full state reset, preserving adaptability while maintaining the admission boundary. Each invariant removes a different unsafe path from the planner to enforcement. I1 prevents malformed or out-of-range intents from entering the compiler path. I2 protects flows in $F_{\mathrm{crit}}$ from restrictive actions unless an explicit policy exemption exists. I3 prevents an admitted action from consuming the headroom reserved for protected services. I4 bounds the rate of policy transitions on a scope and therefore limits adversary-induced oscillation. I5 prevents high-impact actions when the observed severity and uncertainty do not justify them. Section~VI shows how these invariants fire under the three attack classes and under nominal LLM-assisted planning.

\subsection{Integrity Path to Kernel Enforcement}
\label{sec:integrity}

An intent that passes the admission check is not written to enforcement maps directly by the planner or governor. Instead, the governor emits a signed approval record:
\begin{equation}
  \rho_t = \bigl\langle\, t,\; a_t,\; \mathcal{H}(\mathbf{x}_t),\; \mathrm{HMAC}_k\!\bigl(t \,\|\, a_t \,\|\, \mathcal{H}(\mathbf{x}_t)\bigr) \,\bigr\rangle,
  \label{eq:receipt}
\end{equation}
where $\mathcal{H}(\mathbf{x}_t)$ is a hash of the decision-context snapshot and $k$ is held exclusively by the governor process. A privileged compiler service, which is the only component holding eBPF map file descriptors and the relevant Linux capabilities, verifies $\rho_t$ before performing any update. It then translates the approved intent into a bounded set of kernel-facing operations. These include selecting a pre-defined rate tier, enabling a pre-defined inspection path, or writing a schema-validated entry into a policy map.

\subsection{Proposed Split-Control Architecture}
\label{sec:arch}

\begin{figure}[t]
  \centering
  \includegraphics[clip, width=\columnwidth]{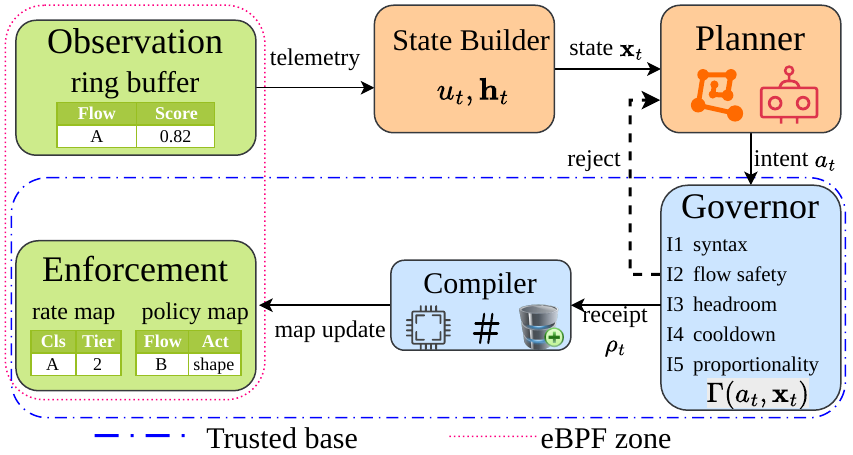}
  \caption{Compact split-control architecture. The planner proposes typed security intents from the built state, but only the governor may admit them for enforcement. Approved actions become bounded eBPF map updates after compiler-side receipt verification. Rejected actions return typed feedback for replanning.}
  \label{fig:arch}
\end{figure}

The proposed framework has six logical components arranged as a unidirectional enforcement pipeline, shown in Fig.~\ref{fig:arch}. Autonomy is confined to the planning stage. All live actuation authority rests within the trusted base.

\noindent\textbf{Observation plane.} Kernel-resident eBPF programs attached at \texttt{tc} and \texttt{XDP} hooks export per-class packet and byte rates, drop counts, queue pressure, protected-flow activity indicators, and policy-map access timestamps through a ring buffer. These programs are pre-compiled and pre-verified at deployment. No user-space component can load or modify them.

\noindent\textbf{State builder.} A user-space daemon assembles $\mathbf{x}_t$ from ring-buffer data, computes $u_t$ as a confidence-weighted function of measurement freshness and cross-source consistency, and maintains $\mathbf{h}_t$ as a bounded sliding window of past governor decisions and observed flow-quality outcomes.

\noindent\textbf{Planner.} The planner receives $\mathbf{x}_t$ and returns a structured intent from the vocabulary in Table~\ref{tab:intents}. The implementation may be an LLM-based agent over a typed tool interface, a learned policy, or a rule-based controller. The architecture is independent of this choice.

\noindent\textbf{Governor.} The governor evaluates $\Gamma(a_t, \mathbf{x}_t)$ per Section~\ref{sec:predicate}, emits $\rho_t$ on admission, and returns a typed rejection with an optional lower-impact suggestion otherwise.

\noindent\textbf{Compiler.} After verifying the HMAC in $\rho_t$, the compiler translates the approved intent into bounded map updates and appends the receipt to the audit log.

\noindent\textbf{Enforcer.} Approved actions become bounded eBPF map updates after compiler-side receipt verification. 

\section{Prototype Implementation}
\label{sec:proto}

\subsection{5G Edge Testbed}

The prototype runs on the University's standalone 5G Test Network (5GTN)~\cite{5gtn2024}, an academic research and experimentation platform operating on the n78 band (3.5 GHz) with an on-premise 5G core. A Raspberry Pi~5 (4-core Cortex-A76, 8 GB RAM, kernel~6.12.79) serves as the enforcement node and gateway. Its \texttt{wlan0} interface connects to the ESP32-S3 endpoint devices over a local WiFi access point, and its \texttt{wwan0} interface provides 5G connectivity to the 5GTN core over the cellular uplink. eBPF programs compiled against a BTF-generated \texttt{vmlinux.h} attach at \texttt{tc} ingress on \texttt{wlan0} and use three BPF maps: an LRU hash for per-source epoch rate counters, a hash for compiler-written rate limits, and a 128 kB ring buffer for telemetry export. All user-space components run on the RPi5, including a Go state builder that assembles $\mathbf{x}_t$ every 100 ms epoch, a Go governor, a privileged Go compiler, and a planner. Fig.~\ref{fig:testbed} shows the testbed topology.

\input{figs/fig_testbed}

Two traffic classes originate from the ESP32-S3 firmware. The \emph{protected} alarm class generates 20-byte UDP datagrams at 10\,Hz, corresponding to a nominal 100\,ms inter-arrival period, while the \emph{routine} class generates 1\,kB telemetry bursts at 1\,Hz. We do not measure the ESP32-to-RPi one-way latency as an evaluation metric because the long runs exposed cross-device clock drift and occasional invalid negative latency samples. Instead, protected-flow service is evaluated using receiver-side cadence. Inter-arrival times are computed directly from RPi receive timestamps, grouped by source IP and sequence number. Baseline runs last approximately 30 minutes, and each attack run lasts approximately five minutes. The LLM planner uses \texttt{llama3.1:8b} in Q4\_K\_M quantized form, served through Ollama~\cite{ollama2026} on a co-located GPU node (RTX~4000 Ada, 20\,GB VRAM). The planner wrapper submits only structured tool-call intents to the governor. Non-tool or no-op outputs from the LLM are logged separately and do not update enforcement state. The prototype uses $\Delta_{\min}=5$ epochs, corresponding to 500\,ms, and $r_{\min}=6.4$\,kbps, a 2$\times$ reserve over the nominal protected-alarm rate.

\subsection{Adversarial Workloads}
The three threat classes in Table~\ref{tab:threats} are executed independently with a 30\,ms cooldown to avoid overlap. For T1, the context-manipulation script suppresses alarm-source reports from two of the three ESP32 nodes before they reach the state builder, causing the planner's view of protected-flow activity to diverge from the receiver-side traffic record. For T2, a test process injects intents directly to the governor's Unix socket at $f_{\mathrm{adv}} = 2/\Delta_{\min}$. This models a co-located adversarial process and represents a conservative worst case, since a remote attacker would face additional transit latency that reduces effective injection frequency. For T3, \texttt{hping3} on the RPi5 injects UDP bursts timed to epoch boundaries, immediately after a governor-approved enforcement transition, to probe the stabilisation window.

\section{Evaluation}
\label{sec:eval}
This section evaluates the proposed split-control design in the 5G test network. The goal is not only to measure enforcement performance but also determine whether the governor preserves the security boundary between planner output and kernel actuation under realistic and adversarial conditions. We therefore examine four aspects: protected-flow cadence, blocking of unsafe actions, control-path overhead, and robustness to planner-side manipulation.

\subsection{Security and Performance Results}
The evaluation uses four configuration labels, listed in Table~\ref{tab:configurations}. We use the friendly labels in figures and prose, while the internal run names are kept in the artifact CSV files for reproducibility. Static is the no-adaptation control, Rule is the deterministic planner with governor, Rule, no gov. is the ablation that removes the admission boundary, and LLM is the proposed planner-governor configuration. \textsc{ReqApproval} and \textsc{Rollback} are part of the vocabulary but are not triggered by the attack workload in this evaluation. Their inclusion lets the proportionality bound contract toward operator-mediated states rather than direct mitigation when uncertainty is high.

\begin{figure}[htpb]
\centering
\includegraphics[width=\columnwidth]{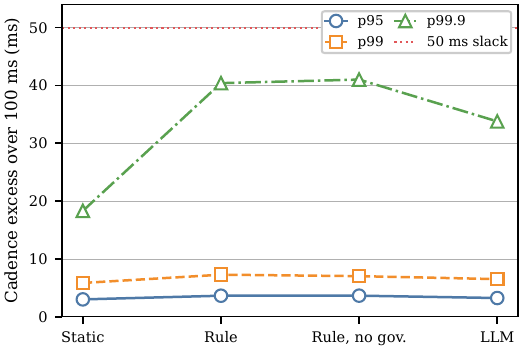}
\vspace{-20pt}
\caption{Protected-flow delivery cadence computed from receiver-side timestamps only. The figure shows excess inter-arrival time over the nominal 100\,ms alarm period. The 50\,ms reference corresponds to the service slack used in the original alarm budget. Sequence-gap counts are reported in Table~\ref{tab:configurations}.}
\label{fig:m1-cadence}
\vspace{-5pt}
\end{figure}

\textbf{M1 -- Receiver-side protected-flow cadence (Fig.~\ref{fig:m1-cadence}, Table~\ref{tab:configurations}).}
The earlier version of the experiment reported ESP32-to-RPi one-way latency. We do not use that metric here because the long runs revealed cross-device clock drift and invalid negative samples. M1 therefore asks a narrower and more reliable question: does the protected alarm stream remain regular at the receiver? Across all four 30-minute baseline runs, the answer is yes. The p99 excess over the nominal 100\,ms alarm period remains below 7.3\,ms, and no sequence gaps are observed in any configuration. The p99.9 excess remains below the 50\,ms service slack for all four baselines. Rare intervals above 150\,ms are reported explicitly in Table~\ref{tab:configurations} rather than hidden inside a latency CDF.

\begin{table}[t]
\centering
\caption{Experimental configurations and receiver-side protected-flow cadence.}
\label{tab:configurations}
\footnotesize
\setlength{\tabcolsep}{2.2pt}
\renewcommand{\arraystretch}{1.02}
\begin{tabularx}{\columnwidth}{@{}p{0.18\columnwidth}Xrrrrr@{}}
\toprule
Config. & Role & $n$ & p99 & p99.9 & gaps & $>150$ \\
        &      & intervals & \multicolumn{2}{c}{excess (ms)} & count & ms \\
\midrule
Static & No adaptation control & 54032 & 5.84 & 18.31 & 0 & 3 \\

Rule & Rule planner with governor & 54092 & 7.27 & 40.40 & 0 & 18 \\

Rule, no gov. & Governor ablation & 54061 & 7.02 & 41.02 & 0 & 23 \\

LLM & LLM planner with governor & 54091 & 6.51 & 33.78 & 0 & 19 \\
\bottomrule
\end{tabularx}

\vspace{2pt}
\begin{minipage}{0.98\columnwidth}
\footnotesize
\emph{Note:} Excess is measured over the nominal 100\,ms alarm period using only RPi receive timestamps.
\end{minipage}
\vspace{-10pt}
\end{table}

\textbf{M2 -- Attack containment (Table~\ref{tab:m2-containment}).}
The attack experiments exercise the governor boundary in three different ways. T1 activates the context-manipulation path in the three-ESP32 testbed: suppressing two alarm sources raises $u_t$ to 0.600 during the attack window. This is not reported as a traffic-latency result. It is evidence that the uncertainty channel responds when the planner's context is manipulated. In T2, I4 rejects 49.7\% of injected oscillation intents, which is the expected alternate-intent clipping pattern at $f_{\mathrm{adv}} = 2/\Delta_{\min}$. T3 remains partially contained as the residual window is bounded but non-zero, with an average window of 2.616\,ms and an average admitted burst of 1279.2\,B. We report this as a bounded residual leak rather than claiming complete elimination of transition exposure. The proportionality contraction engages only when severity also rises; under T1 with nominal traffic, severity stays low and I5 fires only when the LLM proposes high-impact actions in this elevated-uncertainty state. The rule planner under B2 remained at low impact, so its I5 rate stays near zero — see Fig.~\ref{fig:m4-invariants}.

\begin{table}[t]
\centering
\caption{Attack-containment summary.}
\label{tab:m2-containment}
\small
\setlength{\tabcolsep}{3.2pt}
\renewcommand{\arraystretch}{1.02}
\begin{tabularx}{\columnwidth}{@{}lXcl@{}}
\toprule
Attack & Measure & Observed & Result \\
\midrule
T1 & max $u_t$ during attack & 0.600 & YES \\
T2 & I4 rejection rate       & 49.67\% & YES \\
T3 & avg window / burst      & \shortstack{2.616\,ms\\1279\,B} & PARTIAL \\
\bottomrule
\end{tabularx}

\vspace{3pt}
\begin{minipage}{0.98\columnwidth}
\footnotesize
T1: context manipulation; T2: policy oscillation; T3: residual-window attack.
\end{minipage}
% \vspace{-5pt}
\end{table}

\begin{figure}[t]
\centering
\includegraphics[width=\columnwidth]{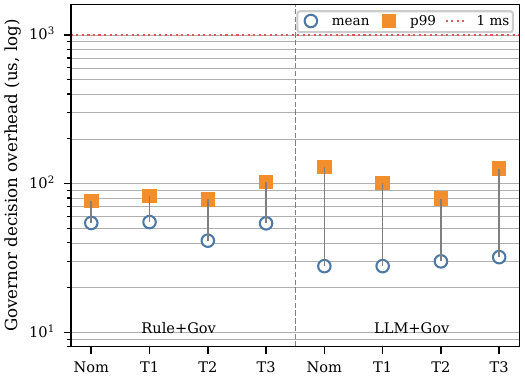}
\vspace{-20pt}
\caption{Governor decision overhead across nominal and attack scenarios. For each scenario, the open circle is the mean and the filled square is the p99; the connecting line is the spread. All p99 values remain below 1\,ms, so the governor is sub-millisecond even under active attack.}
\label{fig:m3-overhead}
% \vspace{-10pt}
\end{figure}

\textbf{M3 -- Governor overhead (Fig.~\ref{fig:m3-overhead}, Table~\ref{tab:m3-overhead}).}
Governor interposition remains sub-millisecond in every scenario. Across nominal and attack runs, p99 decision overhead ranges from 76.24\,$\mu$s to 129.09\,$\mu$s, well below the 1\,ms reference used in Fig.~\ref{fig:m3-overhead}. The result separates the safety boundary from the planner's complexity. The governor check is fast even when the planner is stochastic or the input stream is adversarial. The original 5\,ms engineering guardrail is still comfortably met, but a more impactful result for critical-edge systems is that admission control stays below one millisecond at p99.

\begin{table}[t]
\centering
\caption{Governor decision overhead.}
\label{tab:m3-overhead}
\small
\begin{tabular}{lrrr}
\toprule
Scenario & $n$ & mean ($\mu$s) & p99 ($\mu$s) \\
\midrule
Rule+Gov, nominal & 1801 & 54.34 & 76.24 \\
Rule+Gov, T1 & 301 & 55.11 & 82.56 \\
Rule+Gov, T2 & 601 & 41.33 & 78.52 \\
Rule+Gov, T3 & 321 & 54.08 & 102.22 \\
LLM+Gov, nominal & 51 & 27.86 & 129.09 \\
LLM+Gov, T1 & 10 & 27.86 & 100.78 \\
LLM+Gov, T2 & 309 & 30.00 & 79.20 \\
LLM+Gov, T3 & 25 & 32.03 & 126.26 \\
\bottomrule
\end{tabular}
% \vspace{-5pt}
\end{table}

\begin{figure}[t]
\centering
\includegraphics[width=\columnwidth]{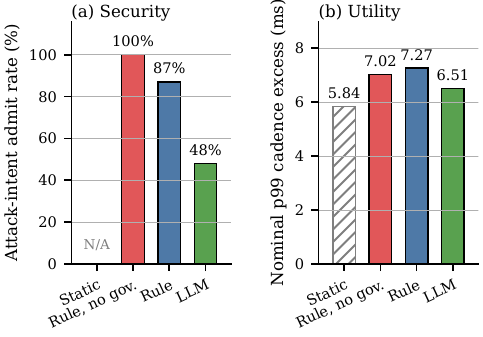}
\caption{Governance tradeoff across four configurations. 
(a) Attack-intent admit rate. Static has no adaptation channel and is marked N/A. Rule, no gov. admits all planner-emitted intents because no admission predicate is present. Rule and LLM with governor admit 86.9\% and 48.0\% of attack-induced structured intents, respectively. 
(b) Nominal p99 cadence excess over the 100\,ms alarm period. All four configurations remain within a 1.5\,ms band. Static is hatched to mark the absence of adaptation.}
\label{fig:governance-tradeoff}
% \vspace{-10pt}
\end{figure}

\begin{figure}[t]
\centering
\includegraphics[width=\columnwidth]{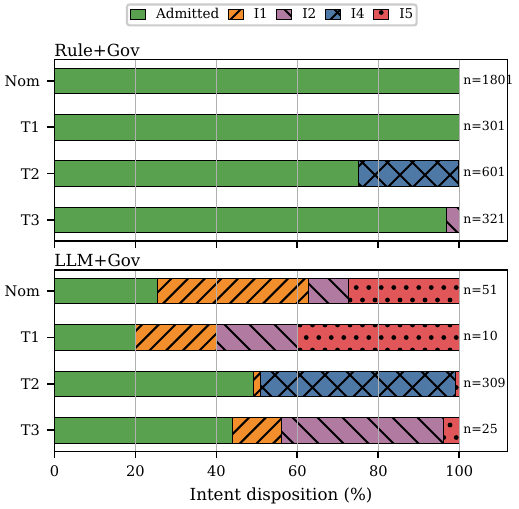}
\vspace{-20pt}
\caption{Intent disposition by governor invariant. T1 represents context manipulation, T2 is policy oscillation, and T3 is residual-window exploitation.}
\label{fig:m4-invariants}
% \vspace{-10pt}
\end{figure}

\textbf{M4 -- Governance tradeoff across configurations (Fig.~\ref{fig:governance-tradeoff}).}
Fig.~\ref{fig:governance-tradeoff} compares the proposed governed configurations with two natural alternatives: Static, which has no adaptation channel, and Rule, no gov., where the rule planner writes to the compiler without the admission predicate. Panel~(a) reports the fraction of attack-induced structured intents that reached enforcement. Static is marked N/A because it emits no adaptive intents. Rule, no gov. admits all planner-emitted intents by construction, while the governed Rule and LLM configurations admit 86.9\% and 48.0\%, respectively, because rejected intents do not reach the compiler. Panel~(b) shows the utility side: nominal p99 cadence excess remains within a 1.5\,ms band across all four configurations. The governor therefore changes what can be actuated under attack without measurably degrading receiver-side protected-flow cadence.

\textbf{M5 -- Planner abuse robustness (Fig.~\ref{fig:m4-invariants}, Table~\ref{tab:llm-examples}).}
The governor's value is clearest when the planner is imperfect. In the nominal LLM run, the wrapper logged 473 LLM decisions. Only 51 were structured intents submitted to the governor; the remaining 422 were non-actuating outputs, mostly prose or JSON-like text converted to \texttt{do\_nothing}. Among the 51 structured intents, 13 were admitted and 38 were rejected before enforcement: 19 by I1, five by I2, and 14 by I5. Table~\ref{tab:llm-examples} gives representative examples. 

Under T2, both planner families show the same I4 signature, with 149 cooldown rejections. Under T3, both paths show ten I2 flow-safety rejections. The same admission boundary applies regardless of how the intent was generated. Invariant I3 reserves $r_{\min}=6.4$\,kbps of bandwidth headroom for the protected class, a 2$\times$ margin over the nominal alarm-class load. In the measured workload, the alarm and routine traffic remain below this floor, so I3 does not activate. This is expected: I3 is a structural guardrail for saturation conditions, while I1, I2, I4, and I5 exercise the planner-governor boundary in the attacks evaluated here.

\begin{table}[t]
\centering
\caption{Examples of structured LLM intents stopped by the governor.}
\label{tab:llm-examples}
\small
\setlength{\tabcolsep}{3pt}
\renewcommand{\arraystretch}{1.03}
\begin{tabularx}{\columnwidth}{@{}p{0.32\columnwidth}p{0.10\columnwidth}X@{}}
\toprule
Proposed intent & Inv. & Governor response \\
\midrule
RATELIMIT with zero rate & I1 &
Invalid parameter; RATELIMIT requires a positive rate. \\
RATELIMIT protected source & I2 &
Restrictive action targets a source carrying alarm-class traffic. \\
ESCALATE global & I5 &
Action impact exceeds the proportionality bound. \\
ISOLATE-NC high-impact scope & I5 &
Isolation impact is not justified by observed severity and uncertainty. \\
\bottomrule
\end{tabularx}
\vspace{-10pt}
\end{table}

\subsection{Deployment Implications}
\label{sec:discussion}

The results apply broadly to mixed-criticality scenarios in edge-could continuum. The protected flow set, the service budget for protected flows, and the cooldown $\Delta_{\min}$ are the main deployment-specific parameters. In a healthcare ward, $\mathcal{F}_{\mathrm{crit}}$ is the alarm class and a manipulated planner that throttles alarms threatens patient safety ~\cite{jointcommission2013alarm,sendelbach2013alarm}. In industrial control, $\mathcal{F}_{\mathrm{crit}}$ becomes the control-plane messages and the service budget is the process cycle time. In vehicular edge, $\Delta_{\min}$ shortens to match faster control decisions. The invariant logic, the HMAC receipt chain, and the eBPF substrate carry over unchanged.

The prototype uses $\Delta_{\min}=5$ epochs (500\,ms) and $r_{\min}=6.4$\,kbps, corresponding to a 2$\times$ reserve over the nominal protected-alarm rate. These parameters expose the expected safety--agility tradeoff. A larger $\Delta_{\min}$ makes oscillation attacks harder by enforcing a longer cooldown between successive policy changes, but also slows adaptation to benign changes in operating conditions. A smaller $\Delta_{\min}$ improves responsiveness, but increases susceptibility to oscillatory manipulation. Likewise, a larger $r_{\min}$ preserves more headroom for protected traffic and tightens the governor's admission boundary, while a smaller $r_{\min}$ increases enforcement flexibility at the cost of a weaker safety margin. We leave a systematic parameter-sensitivity study to future work.

The T2 result reflects why temporal stability matters in clinical and industrial edge settings. An adversary does not need to defeat the enforcement mechanism outright. Repeatedly forcing policy transitions is enough to create short periods of mismatch between the active policy and the traffic condition. I4 clips this behaviour at the admission boundary. In our experiment, 149 of the 300 injected oscillation intents fail the cooldown check, producing the expected near-50\% clipping pattern for $f_{\mathrm{adv}} = 2/\Delta_{\min}$. This gives operators a simple deployment knob. $\Delta_{\min}$ can be chosen with the protected-flow period and acceptable adaptation rate in mind.

The prototype evaluates the planner-governor boundary at the unit-cell scale. Three endpoints and a single enforcement node is the right granularity for the structural claims (I2 flow safety, I4 cooldown bounding, HMAC integrity). Larger-scale evidence possibly with synthetic flow replay across thousands of sources, multi-gateway coordination of cooldown state, and saturation testing for I3 is left for the extended version with a packet-level emulator. 

\subsection{Scope and Implications}

Our prototype evaluates a deliberately small deployment: one RPi5 enforcement node, three endpoints, and a 5G uplink. An open problem is the coordination of governor state across multiple enforcement points in larger deployments.

As importantly, the results suggest that adaptive security does not require trusting the component that makes the decision. The planner can be rule-based, learned, or LLM-assisted, while the same governor independently controls what is allowed to reach the enforcement layer. This separation is useful as planners become more capable but also harder to predict: new planning mechanisms can be introduced without granting them direct control over critical network state. The governor therefore provides a stable security boundary between evolving autonomous intelligence and the enforcement substrate.
The experiments also expose a practical limit of this approach. Admission control can reject unsafe actions and bound how frequently policies change, but it cannot eliminate the short transition window after an action is approved. Applications with stricter timing requirements will therefore need faster enforcement activation in addition to governor-mediated admission.

\section{Conclusion}
\label{sec:conclusion}
Adaptive security in critical edge systems fails not only when a planner misses a threat, but also when enforcement faithfully executes an unsafe decision based on manipulated observations. This paper treats the planner--enforcement boundary as the primary security object and protects it through a governor with five admission invariants. A prototype on the University's 5GTN testbed with in-kernel eBPF enforcement shows that protected alarm delivery remains regular, with p99 cadence excess below 7.3 ms and no sequence gaps across baseline runs. Governor decisions remain sub-millisecond, with p99 below 130 $\mu$s across nominal and attack scenarios. Under policy oscillation, the governor rejects 49.7\% of injected intents, while under nominal LLM planning it rejects 38 of 51 structured intents before enforcement. These results show that adaptive planners can retain flexibility without being granted direct authority over critical enforcement state.

\section*{Acknowledgments}
This research is supported by the Finnish Doctoral Program Network in Artificial Intelligence, AI-DOC (decision number VN/3137/2024-OKM-6), Business Finland funded projects TOMOHEAD (8095/31/2022) and SUNSET-6G (8682/31/2022), and by the Research Council of Finland funded projects 6G Flagship (369116) and Profi6 (336449). The work of Dr. Flavio Esposito is supported by USA NSF Awards CNS \#2133407 and OAC \#2530896.

\bibliographystyle{IEEEtran}
\bibliography{refs}

\end{document}

%% file: figs/fig1.tex
% ── Fig. 1: Threat-annotated pipeline overview ────────────────────────────────
\begin{figure}[htpb]
\centering
\resizebox{\columnwidth}{!}{%
\begin{tikzpicture}[
    font=\small,
    pipe/.style={draw, rounded corners=4pt, minimum width=1.55cm,
                 minimum height=0.88cm, align=center, line width=1.1pt},
    obs/.style  ={pipe, fill=blue!38,   draw=blue!65!black},
    untr/.style ={pipe, fill=orange!42, draw=orange!65!black},
    trust/.style={pipe, fill=teal!38,   draw=teal!65!black},
    arr/.style  ={-Stealth, thick, black},
    darr/.style ={-Stealth, thick, dashed, gray!60!black},
    atkarr/.style={-Stealth, thick, red!70!black, dashed},
    node distance=0.32cm and 0.88cm
  ]
  % ── Pipeline ─────────────────────────────────────────────────────
  \node[obs]   (obs)  {eBPF\\Observer};
  \node[untr,  right=1.25cm of obs]  (sb)   {State\\Builder};
  \node[untr,  right=0.95cm of sb]   (plan) {Planner};
  \node[trust, right=0.95cm of plan] (gov)  {Governor};
  \node[trust, right=0.95cm of gov]  (enf)  {eBPF\\Enforcer};
 
  \draw[arr] (obs)  -- node[above,yshift=2pt,font=\scriptsize\bfseries]{telemetry}   (sb);
  \draw[arr] (sb)   -- node[above,yshift=2pt,font=\scriptsize\bfseries]{$\mathbf{x}_t$} (plan);
  \draw[arr] (plan) -- node[above,yshift=2pt,font=\scriptsize\bfseries]{$a_t$}       (gov);
  \draw[arr] (gov)  -- node[above,yshift=2pt,font=\scriptsize\bfseries]{$\rho_t$}    (enf);
 
  \draw[darr] (gov.south) -- ++(0,-0.82) -|
    node[above,font=\scriptsize\bfseries,pos=0.22]{reject} (plan.south);
 
  % ── Trusted base boundary ────────────────────────────────────────
  \begin{scope}[on background layer]
    \node[draw=teal!65!black, densely dashed, line width=1.3pt,
          fill=teal!7, rounded corners=5pt,
          fit=(gov)(enf), outer sep=0.35cm,
          label={[font=\scriptsize\bfseries,teal!70!black,yshift=-10pt]above:Trusted Base}] {};
  \end{scope}
 
  % ── Adversary (stick figure) ─────────────────────────────────────
  \coordinate (ADV) at ($(sb.south)!0.9!(plan.south) + (0,-1.0)$);
  \filldraw[red!72!black] (ADV) circle (0.19cm);
  \draw[red!72!black,line width=1.3pt] ($(ADV)+(0,-0.19)$) -- ++(0,-0.46);
  \draw[red!72!black,line width=1.3pt] ($(ADV)+(0,-0.34)$) -- ++(-0.30,-0.20);
  \draw[red!72!black,line width=1.3pt] ($(ADV)+(0,-0.34)$) -- ++(0.30,-0.20);
  \draw[red!72!black,line width=1.3pt] ($(ADV)+(0,-0.65)$) -- ++(-0.22,-0.30);
  \draw[red!72!black,line width=1.3pt] ($(ADV)+(0,-0.65)$) -- ++(0.22,-0.30);
  \node[font=\scriptsize\bfseries,red!72!black] at ($(ADV)+(0,-1.12)$) {Adversary};
 
  % ── Attack arrows ────────────────────────────────────────────────
  \draw[atkarr] ($(ADV)+(-0.22,0.19)$) to[out=115,in=250]
    node[midway,left,font=\footnotesize\bfseries,red!70!black,xshift=5pt,yshift=5pt]{\textcircled{\tiny T1}} (obs.south);
  \draw[atkarr] ($(ADV)+(0,0.19)$) to[out=92,in=255]
    node[midway,right,font=\footnotesize\bfseries,red!70!black,xshift=-15pt,yshift=2pt]{\textcircled{\tiny T2}} (plan.south);
  \draw[atkarr] ($(ADV)+(0.22,0.19)$) to[out=65,in=258]
    node[midway,right,font=\footnotesize\bfseries,red!70!black,xshift=10pt,yshift=0pt]{\textcircled{\tiny T3}} (enf.south);
 
  % % ── Legend ───────────────────────────────────────────────────────
  % \node[align=left,font=\tiny,anchor=north west]
  %   at ($(obs.south west)+(0,-2.55)$)
  %   {\textcircled{\tiny 1}\ Context manipulation (telemetry/workload shaping)\\
  %    \textcircled{\tiny 2}\ Oscillation induction (decision-boundary cycling)\\
  %    \textcircled{\tiny 3}\ Residual-window exploitation (transition timing)};
\end{tikzpicture}%
}
\vspace{-20pt}
\caption{Threat-annotated pipeline overview. The adversary targets three interfaces in the adaptation path.}
\label{fig:threats}
\vspace{-10pt}
\end{figure}
 % \textcircled{\tiny T1}~Context manipulation poisons telemetry to mis-steer the planner. \textcircled{\tiny T2}~Oscillation induction alternates inputs to cycle the planner across enforcement states. \textcircled{\tiny T3}~Residual-window exploitation times traffic delivery to coincide with enforcement transitions.

%% file: figs/fig_testbed.tex
% ── Fig: 5GTN Testbed Topology ────────────────────────────────────────────────
\begin{figure}[t]
\centering
\resizebox{\columnwidth}{!}{%
\begin{tikzpicture}[
    font=\small,
    esp/.style  ={draw=orange!70!black, fill=orange!20, rounded corners=3pt,
                  minimum width=1.3cm, minimum height=0.7cm, align=center,
                  line width=0.9pt},
    rpi/.style  ={draw=teal!70!black, fill=teal!18, rounded corners=3pt,
                  minimum width=2.4cm, minimum height=2.6cm, align=center,
                  line width=1.2pt},
    core/.style ={draw=blue!70!black, fill=blue!14, rounded corners=3pt,
                  minimum width=1.5cm, minimum height=0.7cm, align=center,
                  line width=0.9pt},
    lbl/.style  ={font=\scriptsize, align=center},
    arr/.style  ={-Stealth, thick},
    warr/.style ={-Stealth, thick, orange!60!black},
    garr/.style ={-Stealth, thick, blue!60!black},
    node distance=0.5cm
  ]

  %% ── ESP32 devices (left column) — lowered by 0.2 to align with RPi ──────
  \node[esp] (e1) at (0, 2.0)   {ESP32\\(alarm)};
  \node[esp] (e2) at (0, 0.9)   {ESP32\\(alarm)};
  \node[esp] (e3) at (0, -0.2)  {ESP32\\(routine)};

  %% ── RPi5 enforcement node — centre lowered from 1.1 to 0.9 ──────────────
  %% Box now spans y = -0.4 to 2.2; title labels at 2.35 and 2.65 sit above
  \node[rpi] (rpi) at (3.8, 0.9) {};

  %% Title above the box (now visible — both above box top y=2.2)
  \node[font=\small\bfseries, teal!70!black]    at (3.8, 2.65) {RPi~5};
  \node[font=\scriptsize\itshape, teal!60!black] at (3.8, 2.35) {enforcement node};

  %% Software stack labels (shifted down by 0.2 to stay centred in box)
  \node[font=\scriptsize, align=center] at (3.8, 1.55)
    {\textbf{eBPF hooks}\\(\texttt{tc/wlan0})};
  \node[font=\scriptsize] at (3.8, 0.85) {State Builder};
  \node[font=\scriptsize] at (3.8, 0.45) {Governor};
  \node[font=\scriptsize] at (3.8, 0.05) {Planner};

  %% ── 5GTN core (right) ────────────────────────────────────────────────────
  \node[core] (upf) at (7.6, 1.85) {UPF};
  \node[core] (amf) at (7.6, 0.95) {AMF/SMF};
  \node[font=\small\bfseries, blue!70!black]     at (7.6, 2.65) {5GTN Core};
  \node[font=\scriptsize\itshape, blue!60!black]  at (7.6, 2.35) {(n78 / 3.5\,GHz)};

  %% ── WiFi links (ESP32 → RPi left edge at each device's y) ──────────────
  \draw[warr] (e1.east) -- (rpi.west |- e1);
  \draw[warr] (e2.east) -- (rpi.west |- e2);
  \draw[warr] (e3.east) -- (rpi.west |- e3);

  %% WiFi / wlan0 group label (shifted down to track ESP32 cluster)
  \node[lbl, orange!70!black]                    at (1.55, 2.52) {WiFi};
  \node[lbl, orange!70!black, font=\scriptsize]  at (1.55, 2.25) {(wlan0)};

  %% ── 5G uplink (RPi right edge → UPF/AMF) ────────────────────────────────
  \draw[garr] (rpi.east |- upf) -- node[above, lbl]{5G uplink} (upf.west);
  \draw[garr, dashed] (rpi.east |- amf) -- (amf.west);

  %% wwan0 label
  \node[lbl, blue!60!black, font=\scriptsize]  at (5.7, 2.40) {(wwan0)};

  %% ── Trusted-base boundary (extended down to enclose lowered box) ────────
  \begin{scope}[on background layer]
    \draw[densely dashed, teal!55, line width=1.1pt, rounded corners=4pt]
      (1.85, -0.50) rectangle (5.75, 2.85);
  \end{scope}

  %% ── Legend (lowered to clear the new boundary bottom) ───────────────────
  \node[lbl, align=left, font=\tiny] at (3.8, -0.85)
    {\textcolor{orange!70!black}{$\blacksquare$}~Untrusted endpoint \quad
     \textcolor{teal!70!black}{$\square$}~Trusted base (RPi5) \quad
     \textcolor{blue!70!black}{$\blacksquare$}~5GTN core};

\end{tikzpicture}%
}
\caption{Testbed topology. ESP32 nodes connect to the RPi~5 enforcement node over WiFi (\texttt{wlan0}). The RPi~5 runs the state builder, governor, and planner, with eBPF hooks on \texttt{wlan0} for enforcement. The \texttt{wwan0} interface provides 5G connectivity to the 5GTN core.}
\label{fig:testbed}
\vspace{-10pt}
\end{figure}

%% file: refs.bib
@inproceedings{sangiorgi2025deauth,
  author    = {Sangiorgi, A. and Pinto, A. and Tourani, R. and Esposito, F.},
  title     = {Mitigating De-authentication {DoS} Attacks in 802.11 via {eBPF} and {XDP}},
  booktitle = {Proceedings of the IEEE Conference on Network Softwarization (NetSoft)},
  year      = {2025},
  address   = {Budapest, Hungary},
  month     = {June},
  pages     = {}
}

@misc{nist2026rfi,
  author       = {{National Institute of Standards and Technology}},
  title        = {{CAISI} Issues Request for Information About Securing
                  {AI} Agent Systems},
  year         = {2026},
  month        = jan,
  note         = {NIST news release, January~12, 2026},
  url          = {https://www.nist.gov/news-events/news/2026/01/caisi-issues-request-information-about-securing-ai-agent-systems}
}

@misc{nist2026monitoring,
  author       = {{National Institute of Standards and Technology}},
  title        = {Challenges to the Monitoring of Deployed {AI} Systems},
  howpublished = {NIST AI 800-4},
  year         = {2026},
  month        = mar,
  url          = {https://nvlpubs.nist.gov/nistpubs/ai/NIST.AI.800-4.pdf}
}

@inproceedings{greshake2023indirect,
  author    = {Kai Greshake and Sahar Abdelnabi and Shailesh Mishra and
               Christoph Endres and Thorsten Holz and Mario Fritz},
  title     = {Not What You've Signed Up For: Compromising Real-World
               {LLM}-Integrated Applications with Indirect Prompt Injection},
  booktitle = {Proceedings of the 16th {ACM} Workshop on Artificial
               Intelligence and Security (AISec~'23)},
  year      = {2023},
  doi       = {10.1145/3605764.3623985}
}

@article{chatzimiltis2025agentic,
  title={Agentic {AI} for {6G}: A New Paradigm for Autonomous RAN Security Compliance},
  author={Chatzimiltis, Sotiris and Mashhadi, Mahdi Boloursaz and Shojafar, Mohammad and Debbah, Merouane and Tafazolli, Rahim},
  journal={arXiv preprint arXiv:2512.12400},
  year={2025}
}

@INPROCEEDINGS{sharma2025mobillm,
  author={Wen, Haohuang and Sharma, Prakhar and Yegneswaran, Vinod and Gehani, Ashish and Porras, Phillip and Lin, Zhiqiang},
  booktitle={MILCOM 2025 - 2025 IEEE Military Communications Conference (MILCOM)}, 
  title={MobiLLM: An Agentic AI Framework for Closed-Loop Threat Mitigation in 6G Open RANs}, 
  year={2025},
  volume={},
  number={},
  pages={1-6},
  doi={10.1109/MILCOM64451.2025.11310651}}

@techreport{etsi_zsm009,
  author      = {{ETSI}},
  title       = {Zero-touch network and Service Management ({ZSM});
                 Closed-Loop Automation},
  institution = {European Telecommunications Standards Institute},
  number      = {GS ZSM 009-1 v1.1.1},
  year        = {2021},
  url         = {https://www.etsi.org/deliver/etsi_gs/ZSM/001_099/00901/01.01.01_60/gs_zsm00901v010101p.pdf}
}

@ARTICLE{ahmad2025adapt,
  author={Ahmad, Ijaz and Gimhana, Shakthi and Ahmad, Ijaz and Harjula, Erkki},
  journal={IEEE Networking Letters}, 
  title={Adaptive Trust Architecture for Secure {IoT} Communication in {6G}}, 
  year={2025},
  volume={7},
  number={2},
  pages={113-116},
  doi={10.1109/LNET.2025.3566909}}

@misc{rfc9315,
  author       = {A. Clemm and L. Ciavaglia and
                  L.~Z. Granville and J. Tantsura},
  title        = {Intent-Based Networking---Concepts and Definitions},
  howpublished = {RFC~9315},
  publisher    = {IETF},
  year         = {2022},
  doi          = {10.17487/RFC9315}
}

@techreport{oran_wg11_threat,
  author      = {{O-RAN ALLIANCE}},
  title       = {{O-RAN} Security Threat Modeling and Risk Assessment},
  institution = {O-RAN ALLIANCE},
  type        = {Technical Report},
  number      = {O-RAN.WG11.Threat-Model},
  year        = {2024},
  url         = {https://www.o-ran.org/specifications}
}

@ARTICLE{oRanSecurity2025,
  author={Alam, Khurshid and Habibi, Mohammad Asif and Tammen, Matthias and Krummacker, Dennis and Saad, Walid and Renzo, Marco Di and Melodia, Tommaso and Costa-Pérez, Xavier and Debbah, Mérouane and Dutta, Ashutosh and Schotten, Hans D.},
  journal={IEEE Communications Surveys \& Tutorials}, 
  title={A Comprehensive Tutorial and Survey of O-RAN: Exploring Slicing-Aware Architecture, Deployment Options, Use Cases, and Challenges}, 
  year={2026},
  volume={28},
  number={},
  pages={1637-1678},
  doi={10.1109/COMST.2025.3598406}}

@article{fue2025private5g,
  title={Understanding Security Vulnerabilities in Private 5G Networks: Insights from a Literature Review},
  author={Fue, Jacinta and Gutierrez, Jairo A and Donoso, Yezid},
  journal={Future Internet},
  volume={17},
  number={11},
  pages={485},
  year={2025},
  publisher={MDPI}
}

@inproceedings{jin2024beebox,
  author    = {Di Jin and Alexander J. Gaidis and Vasileios P. Kemerlis},
  title     = {{BeeBox}: Hardening {BPF} against Transient Execution Attacks},
  booktitle = {33rd {USENIX} Security Symposium (USENIX Security~'24)},
  year      = {2024}
}

@inproceedings{sun2025aee,
  author    = {Hao Sun and Zhendong Su},
  title     = {Approximation Enforced Execution of Untrusted
               {Linux} Kernel Extensions},
  booktitle = {34th {USENIX} Security Symposium (USENIX Security~'25)},
  year      = {2025}
}

@inproceedings{jia2025rex,
  author    = {Jinghao Jia and Ruowen Qin and Milo Craun and
               Egor Lukiyanov and Ayush Bansal and Minh Phan and
               Tianyin Xu},
  title     = {Rex: Closing the Language-Verifier Gap with Safe and
               Usable Kernel Extensions},
  booktitle = {2025 {USENIX} Annual Technical Conference (USENIX ATC~'25)},
  year      = {2025}
}

@inproceedings{zheng2025agentsight,
  author    = {Yusheng Zheng and Yanpeng Hu and Tong Yu and Andi Quinn},
  title     = {{AgentSight}: System-Level Observability for {AI} Agents
               Using {eBPF}},
  booktitle = {Proceedings of {SOSP}},
  year      = {2025},
  doi       = {10.1145/3766882.3767169}
}

@misc{cilium2026intro,
  author       = {{Cilium Authors}},
  title        = {Introduction to Cilium and Hubble},
  year         = {2026},
  note         = {Project documentation},
  url          = {https://docs.cilium.io/en/stable/overview/intro/}
}

@ARTICLE{soldani2023ebpf,
  author={Soldani, David and Nahi, Petrit and Bour, Hami and Jafarizadeh, Saber and Soliman, Mohammed F. and Di Giovanna, Leonardo and Monaco, Francesco and Ognibene, Giuseppe and Risso, Fulvio},
  journal={IEEE Access}, 
  title={eBPF: A New Approach to Cloud-Native Observability, Networking and Security for Current (5G) and Future Mobile Networks (6G and Beyond)}, 
  year={2023},
  volume={11},
  number={},
  pages={57174-57202},
  doi={10.1109/ACCESS.2023.3281480}}

@misc{5gtn2024,
  author       = {{University of Oulu}},
  title        = {{5GTN}: 5G Test Network},
  year         = {2024},
  howpublished = {University of Oulu research infrastructure},
  url          = {https://5gtn.fi}
}

@techreport{anderson1972computer,
  author      = {James P. Anderson},
  title       = {Computer Security Technology Planning Study},
  institution = {Electronic Systems Division, Air Force Systems Command},
  number      = {ESD-TR-73-51},
  year        = {1972},
  note        = {The foundational reference monitor concept}
}

@INPROCEEDINGS{gimhana2025mmtc,
  author={Gimhana, Shakthi and Ahmad, Ijaz and Porambage, Pawani and Harjula, Erkki},
  booktitle={2025 Joint European Conference on Networks and Communications \& 6G Summit (EuCNC/6G Summit)}, 
  title={Mitigating {DoS} Attacks in {mMTC}: An Energy Efficiency Perspective}, 
  year={2025},
  volume={},
  number={},
  pages={199-204},
  doi={10.1109/EuCNC/6GSummit63408.2025.11036916}}

@article{jointcommission2013alarm,
  author       = {{The Joint Commission}},
  title        = {Medical Device Alarm Safety in Hospitals},
  journal      = {Sentinel Event Alert},
  number       = {50},
  pages        = {1--3},
  year         = {2013},
  month        = apr,
  note         = {PMID: 23767076},
  url          = {https://www.jointcommission.org/en-us/knowledge-library/newsletters/sentinel-event-alert/issue-50}
}

@article{sendelbach2013alarm,
  author  = {Sendelbach, Sue and Funk, Marjorie},
  title   = {Alarm Fatigue: {A} Patient Safety Concern},
  journal = {AACN Advanced Critical Care},
  volume  = {24},
  number  = {4},
  pages   = {378--386},
  year    = {2013},
  doi     = {10.4037/NCI.0b013e3182a903f9}
}

@article{schneider2000enforceable,
author = {Schneider, Fred B.},
title = {Enforceable security policies},
year = {2000},
issue_date = {Feb. 2000},
publisher = {Association for Computing Machinery},
address = {New York, NY, USA},
volume = {3},
number = {1},
issn = {1094-9224},
url = {https://doi.org/10.1145/353323.353382},
doi = {10.1145/353323.353382},
journal = {ACM Trans. Inf. Syst. Secur.},
month = feb,
pages = {30–50},
numpages = {21}
}

@article{sanchez2019runtime,
  title={A survey of challenges for runtime verification from advanced application domains (beyond software)},
  author={S{\'a}nchez, C{\'e}sar and Schneider, Gerardo and Ahrendt, Wolfgang and Bartocci, Ezio and Bianculli, Domenico and Colombo, Christian and Falcone, Yli{\`e}s and Francalanza, Adrian and Krsti{\'c}, Sr{\dj}an and Louren{\c{c}}o, Jo{\~a}o M and others},
  journal={Formal Methods in System Design},
  volume={54},
  number={3},
  pages={279--335},
  year={2019},
  publisher={Springer}
}

@inproceedings{alshiekh2018shielding,
  title={Safe reinforcement learning via shielding},
  author={Alshiekh, Mohammed and Bloem, Roderick and Ehlers, R{\"u}diger and K{\"o}nighofer, Bettina and Niekum, Scott and Topcu, Ufuk},
  booktitle={Proceedings of the AAAI conference on artificial intelligence},
  volume={32},
  number={1},
  year={2018}
}

@article{cardenas2008control,
  title={Research challenges for the security of control systems.},
  author={C{\'a}rdenas, Alvaro A and Amin, Saurabh and Sastry, Shankar},
  journal={HotSec},
  volume={5},
  number={15},
  pages={1158},
  year={2008},
  publisher={CA, United States}
}

@INPROCEEDINGS{ahmad24adaptive,
  author={Ahmad, Ijaz and Shahid, Faheem and Ahmad, Ijaz and Islam, Johirul and Haque, Kazi Nymul and Harjula, Erkki},
  booktitle={2024 18th International Symposium on Medical Information and Communication Technology (ISMICT)}, 
  title={Adaptive Lightweight Security for Performance Efficiency in Critical Healthcare Monitoring}, 
  year={2024},
  volume={},
  number={},
  pages={78-83},
  doi={10.1109/ISMICT61996.2024.10738175}}

@article{liu2021vec,
  title={Vehicular edge computing and networking: A survey},
  author={Liu, Lei and Chen, Chen and Pei, Qingqi and Maharjan, Sabita and Zhang, Yan},
  journal={Mobile networks and applications},
  volume={26},
  number={3},
  pages={1145--1168},
  year={2021},
  publisher={Springer}
}

@inproceedings{ahmad2026adaptive,
  title={Adaptive Security at the Edge for {6G}-Enabled Healthcare {IoT}},
  booktitle={2026 Joint European Conference on Networks and Communications \& 6G Summit (EuCNC/6G Summit)},
  author={Ahmad, Ijaz and Ahmad, Ijaz and Harjula, Erkki},
  doi={10.48550/arXiv.2607.27858},
  year={2026}
}

@misc{ollama2026,
  author       = {{Ollama}},
  title        = {Ollama},
  year         = {2026},
  version      = {X.Y.Z},
  url          = {https://github.com/ollama/ollama},
  note         = {Local large language model serving framework. Accessed 2026-05-16}
}
